\documentclass[aps,prl,reprint,superscriptaddress,showpacs,longbibliography,floatfix,footnoteinbib]{revtex4-1}
\usepackage[latin9]{inputenc}
\usepackage{color}
\usepackage{bm}
\usepackage{amsmath}
\usepackage{amssymb}
\usepackage{graphicx}
\usepackage{wasysym}
\usepackage[unicode=true,
 bookmarks=false,
 breaklinks=false,pdfborder={0 0 1},backref=false,colorlinks=true]
 {hyperref}
\hypersetup{pdftitle={Title},
 plainpages=false,pdfpagelabels,linkcolor=red,urlcolor=blue,citecolor=blue,pdfdisplaydoctitle=true,pdfduplex=DuplexFlipLongEdge}

\makeatletter
\usepackage{units}\usepackage{wasysym}

\newcommand{\beginsupplement}{%
        \setcounter{table}{0}
        \renewcommand{\thetable}{S\arabic{table}}%
        \setcounter{figure}{0}
        \renewcommand{\thefigure}{S\arabic{figure}}%
}

\usepackage{bm}
\usepackage{braket}

\makeatother

\begin{document}

\title{Temperature-driven gapless topological insulator}

\author{Miguel Gonçalves}

\affiliation{CeFEMA, Instituto Superior Técnico, Universidade de Lisboa, Av. Rovisco
Pais, 1049-001 Lisboa, Portugal}

\author{Pedro Ribeiro}

\affiliation{CeFEMA, Instituto Superior Técnico, Universidade de Lisboa, Av. Rovisco
Pais, 1049-001 Lisboa, Portugal}

\affiliation{Beijing Computational Science Research Center, Beijing 100084, China}

\author{Rubem Mondaini}

\affiliation{Beijing Computational Science Research Center, Beijing 100084, China}

\author{Eduardo V. Castro}

\affiliation{CeFEMA, Instituto Superior Técnico, Universidade de Lisboa, Av. Rovisco
Pais, 1049-001 Lisboa, Portugal}

\affiliation{Beijing Computational Science Research Center, Beijing 100084, China}

\affiliation{Centro de Física das Universidades do Minho e Porto, Departamento
de Física e Astronomia, Faculdade de Ciências, Universidade do Porto,
4169-007 Porto, Portugal}
\begin{abstract}
We investigate the phase diagram of the Haldane-Falicov-Kimball model
-- a model combining topology, interactions and spontaneous disorder
at finite temperatures. Using an unbiased numerical method, we map
out the phase diagram on the interaction--temperature plane. Along
with known phases, we unveil an \textit{insulating charge ordered
state with gapless excitations }and a temperature-driven \textit{gapless
topological insulating} phase. Intrinsic -- temperature generated
-- disorder, is the key ingredient explaining the unexpected behavior.
Our findings support the possibility of having temperature-driven
topological phase transitions into gapped and gapless topological
insulating phases in systems with a large mass unbalance in fermionic
species. 
\end{abstract}
\maketitle
Understanding the effects of disorder, interactions and temperature
on topological phases of matter is essential to predict the topological
properties and their stability in real-world materials \citep{Xu2006}.
Some of these effects are quite subtle and may have dichotomic features.
For example, topological phases are suppressed in the presence of
strong nearest-neighbor (NN) \citep{Varney2010} or Hubbard-like interactions
\citep{Rachel2010, Yamaji2011, Zheng2011, Yu2011, Griset2012, Hohenadler2011, Hohenadler2012, Reuther2012, Araujo2013, Laubach2014}.
However, interaction-induced magnetic order was found to coexist with
topological phases \citep{Mong2010, Fang2013, Yoshida2013, Miyakoshi2013}
and some studies showed that interactions themselves could induce
a topological phase on a trivial band, forming the so-called topological
Mott insulator \citep{Raghu2008, Wen2010, Budich2012, Dauphin2012, Weeks2010, LeiWang2012, Ruegg2011, Yang2011}.
Even if this phase is disputed outside the mean-field scope in
some models \citep{Garcia_Martinez2013, Daghofer2014, Motruk2015, Capponi2015},
it has been confirmed in others \citep{Budich2013,Rachel2018}.

\begin{figure}[b!]
\centering{}\includegraphics[width=1\columnwidth]{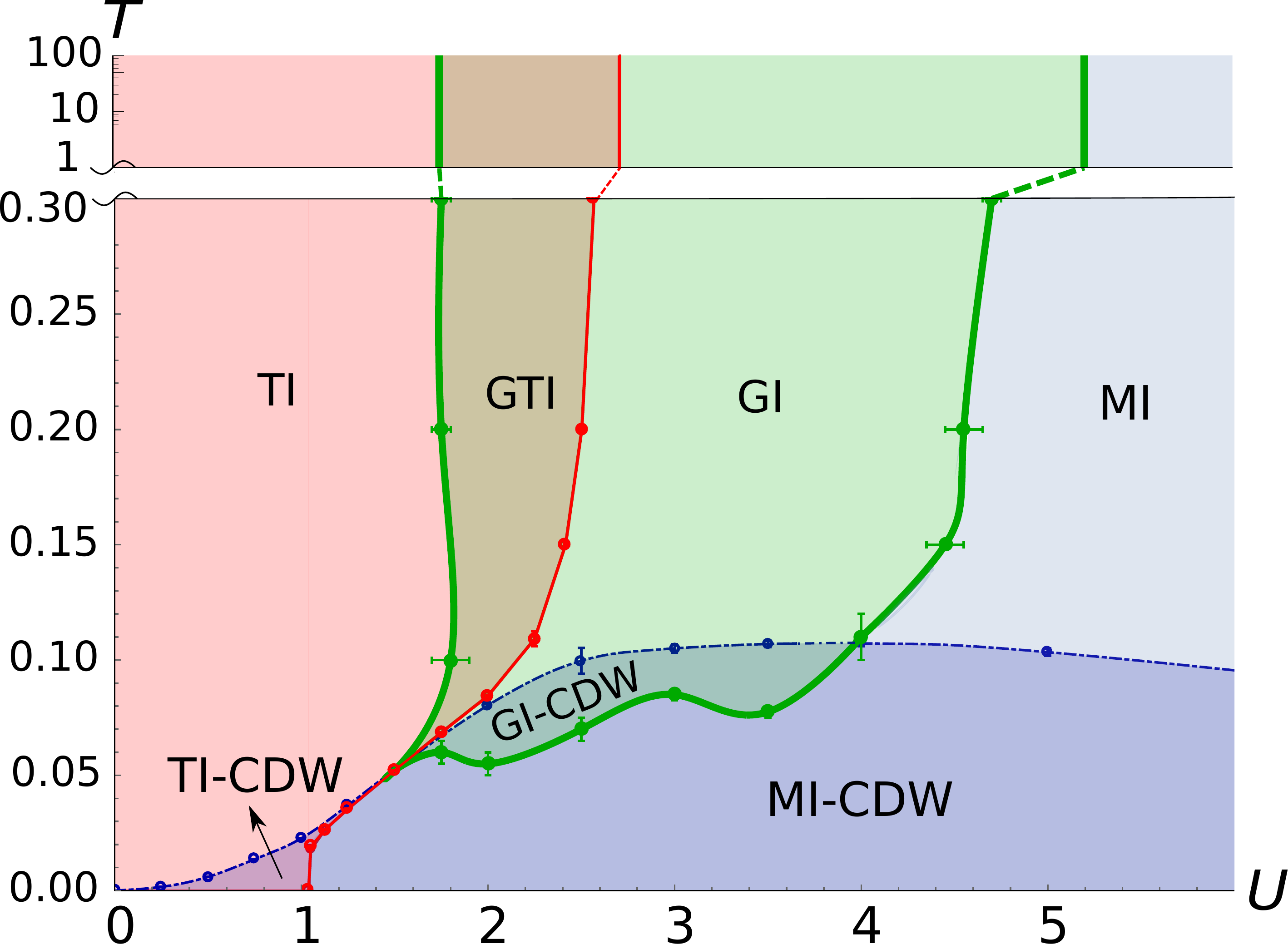}\caption{Phase diagram of the HFKM in the interaction -- temperature plane
obtained with the Monte Carlo method. Phases at intermediate- to high-$T$,
outside the charge density wave phase (CDW): topological insulator
(TI) for small $U$, gapless topological insulator (GTI) and gapless
insulator (GI) for intermediary $U$, and Mott-like insulating phase
(MI) for large $U$. Phases at low-$T$, inside the CDW phase: phases
with similar features as their high-$T$ counterparts were found and
the suffix ``-CDW'' was added. The thin (red) and dashed-dotted
(blue) curves correspond, respectively, to the topological and CDW
phase transitions and the thick (green) curve bounds the gapless region
of the phase diagram. \label{fig:PhaseDiagram_HFKM_Main}}
\end{figure}

The influence of correlations at finite temperatures on topological
insulators (TI) also shows opposite trends \citep{Zhu2014, Chen2015, Zdulski2015}.
Although thermal fluctuations are responsible for the destruction
of topological order when large enough \citep{Yoshida2012, Rivas2013},
they can also drive different types of topological phases \citep{Zhu2014, Dziawa2012}.

The role of disorder on topological phases is also subtle. For TI
within the unitary class \citep{AZ97,evers-at2008,Chiu2016} (for
which time-reversal symmetry is broken), disorder effects localize
every eigenstate except two bulk extended states that carry opposite
Chern numbers \citep{Onoda2003, Onoda2007}. The merging of these
states, for a sufficiently large disorder strength, is associated
with the destruction of the topological phase. Interestingly, a disorder-induced
transition into a new topologically nontrivial phase -- the topological
Anderson insulator (TAI) -- was also shown to be possible \citep{Li2009, Groth2009, Song2012, Su2016, Orth2016}.

In this Letter, we explore some of the dichotomic aspects above by
fully characterizing an interacting quantum model that crucially combines
non-trivial topology, disorder, temperature, and interaction effects,
and which can be efficiently studied by unbiased numerical methods.
Our main results are summarized in Fig.~\ref{fig:PhaseDiagram_HFKM_Main},
which depicts the different phases as a function of the temperature,
$T$, and of the interaction magnitude, $U$. As a central result,
topological order was found to appear for intermediate $U$ values
when $T$ is increased, and to extend into the gapless region of the
phase diagram at higher $T$, giving rise to a \emph{temperature-driven
gapless topological insulating phase} (GTI).

\paragraph{Model.---}

Our starting point is the Falicov-Kimball model (FKM) \citep{Falicov1969},
a limiting case of the Hubbard model for which one of the spin fermion
species is infinitely massive, rendering these fermions -- the $f$-electrons
-- immobile. For a half-filled bipartite lattice at $T=0$, the $f$-electrons
order in a charge density wave (CDW) state for any finite value of
the interaction strength between the localized and itinerant electrons
\citep{Brandt1986, Kennedy1986, Maska2006, Zonda2009}. Recently,
the full $T$-dependent phase diagram of the 2D FKM was obtained with
Monte Carlo (MC) techniques unveiling an Anderson insulating phase
overlooked in previous studies \citep{Antipov2016}. The averaging
on the configurations of $f$-electrons, sampled according to the
total partition function with the MC method, acts as a disorder potential
to itinerant electrons rendering possible that their eigenstates become
localized without the presence of explicit disorder. We combine the
interacting nature of the FKM with the topological features of the
first theoretical model of a TI under a zero net magnetic field -
the Haldane model \citep{Haldane1988} - which, although robust to
small disorder, has its topological properties destroyed for large
enough disorder strengths \citep{Sheng2005, Zhang2013, EVCastro2015, EVCastro2016}.

\begin{figure}
\centering{}\includegraphics[width=1\columnwidth]{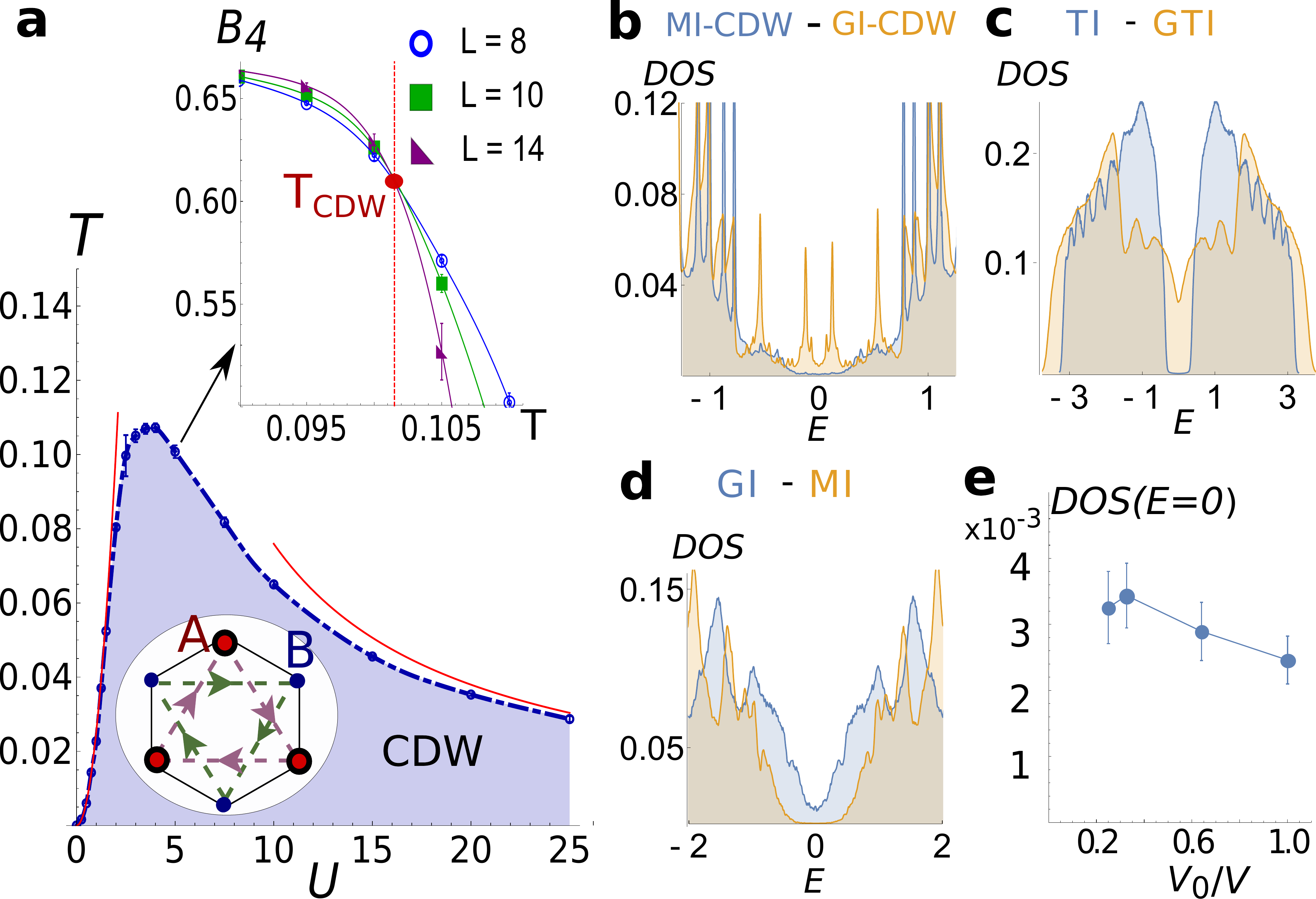}\caption{\textbf{a}, Monte Carlo results of the CDW phase transition along
with the small and large $U$ curves obtained with second order perturbation
theory~\citep{SM}. The larger sites in the honeycomb unit cell inside
the CDW phase represent occupied sites indicating a checkerboard order;
the arrows represent the flow of NNN hoppings. The inset shows an
example of the usage of the Binder cumulant method to compute the
critical temperature $T_{\mathrm{CDW}}$ that corresponds to the intersection
of the obtained curves for different system sizes, for a case where
$U=5$. \textbf{b-d}, Density of states for different points in the
phase diagram: MI-CDW-$(U,T)=(2.5,0.045)$, GI-CDW-$(2.5,0.085)$,
TI-$(1,0.1)$, GTI-$(2,0.1)$ , GI-$(4,0.1)$ and MI-$(5,0.1)$; we
use a Lorentzian broadening of width 0.01. \textbf{e}, Finite-size
scaling of the DOS at $E=0$ for the point $(2.5,0.085)$ used in
\textbf{b}. $V_{0}$ corresponds to volume of the smallest lattice
used (with $L=8$); the DOS($E=0$) was computed in an energy window
corresponding to $1\%$ of the full bandwidth for the $L=8$ system.
This window was reduced proportionally to the system size for larger
systems. \label{fig:CDW_DOS_panel} }
\end{figure}

The Hamiltonian of the Haldane-Falicov-Kimball model (HFKM) is defined
as 
\begin{equation}
\begin{aligned}\hat{H} & =-t\sum_{\langle i,j\rangle}\hat{c}_{i}^{\dagger}\hat{c}_{j}+{\rm i}t_{2}\sum_{\langle\langle i,j\rangle\rangle}\nu_{ij}\hat{c}_{i}^{\dagger}\hat{c}_{j}+{\rm H.c.}\\
 & +U\sum_{i}\hat{c}_{i}^{\dagger}\hat{c}_{i}n_{i}^{f}-\sum_{i}(\mu_{c}\hat{c}_{i}^{\dagger}\hat{c}_{i}+\mu_{f}n_{i}^{f}),
\end{aligned}
\end{equation}
depicting a species of itinerant electrons ($c$-electrons) with creation
operators $\hat{c}_{i}^{\dagger}$ and another of localized electrons
($f$-electrons) whose local density at site $i$ is given by the
number $n_{i}^{f}$. The operators $\hat{c}_{i}^{\dagger}=\hat{c}_{i,A}^{\dagger},\hat{c}_{i,B}^{\dagger}$
are defined in the two interpenetrating triangular sublattices \textbf{$A$}
and \textbf{$B$} that form the honeycomb lattice shown in the sketch
in Fig.\,\ref{fig:CDW_DOS_panel}~\textbf{a}, with total volume
$V=2L^{2}$, where $L$ indicates the linear number of unit cells.
The first term is the kinetic energy of the itinerant electrons associated
with NN hoppings, with $t$ being the hopping integral for NN. The
second term considers Haldane next-nearest neighbor (NNN) complex
hoppings with $\nu_{ij}=\pm1$, according to the arrows represented
in the honeycomb cell in Fig.\,\ref{fig:CDW_DOS_panel}~\textbf{a}.
The third term describes the local interaction between localized and
itinerant electrons, with $U>0$. The final term contains the chemical
potentials for the itinerant and localized electrons, respectively
$\mu_{c}$ and $\mu_{f}$. We focus on the case of half-filling for
both species -- one particle per unit cell -- and therefore set
$\mu_{c}=\mu_{f}=U/2$. In what follows, $t=1$ sets the energy scale
and $t_{2}=0.1t$.

\noindent 

Given that the $f$-electron densities $n_{i}^{f}$ ($=0,1$) can
be seen as classical variables, the partition function can be written
as 
\begin{equation}
\mathcal{Z}=\sum_{\{n_{f}\}}{\rm Tr}_{c}[e^{-\beta\hat{H}(\{n_{f}\})}]=\sum_{\{n_{f}\}}e^{-\beta\mathcal{H}(\{n_{f}\})},\label{eq:partition_func}
\end{equation}

\noindent where

\begin{equation}
\mathcal{H}(\{n_{f}\})=-\frac{U}{2}\sum_{i}n_{i}^{f}-\frac{1}{\beta}\sum_{j}\ln[1+e^{-\beta E_{j}(\{n_{f}\})}]\label{eq:effective_hamil}
\end{equation}

\noindent is the effective Hamiltonian obtained after taking the trace
over the $c$-electrons' degrees of freedom and is defined in terms
of the eigenvalues $E_{j}(\{n_{f}\})$ of $\hat{H}$ obtained for
a fixed configuration $\{n_{f}\}$. In this from, the model is amenable
to classical Monte Carlo sampling.

\paragraph{Observables.---}

For the $f$-electrons, we focus on describing the CDW phase transition,
characterized by an order parameter corresponding to the staggered
occupation of $f$-electrons on sublattices $A$ and $B$, $n_{{\rm st.}}^{f}=n_{A}^{f}-n_{B}^{f}$,
where $n_{x}^{f}$ is the density in sublattice $x$. The critical
$T$ curve, $T_{{\rm {\rm CDW}}}(U)$, is obtained by fixing $U$
and computing the intersections of the $T$-dependent Binder cumulant,
$B_{4}=(1-\langle{n_{{\rm st.}}^{f}}^{4}\rangle/3{\langle n_{{\rm st.}}^{f}}^{2}\rangle^{2})$,
for different system sizes as shown in the inset of Fig.\,\ref{fig:CDW_DOS_panel}~\textbf{a}.
Regarding the $c$-electrons, we investigate the following observables:
the Chern number $C$, computed with the method introduced in Ref.\,\citep{Zhang2013},
specially developed for systems that are not translationally invariant
(see discussion for the validity of this approach below); density
of states (DOS), obtained with the eigenvalues of the fermionic degrees
of freedom; localization of the eigenstates, studied with the inverse
participation ratio (IPR) and level spacing statistics (LSS) methods.
In our definition, the IPR is computed for a given energy $E_{\alpha}$
as ${\cal I}_{\alpha}=\sum_{i}|\phi_{i}^{\alpha}|^{4}$, where $\phi_{i}^{\alpha}$
is the amplitude of the eigenvector with energy $E_{\alpha}$ at site
$i$. An IPR histogram can then be obtained as a function of the energy
by sampling a large number of MC configurations; this quantity scales
to zero with the system's volume if we deal with extended states,
and to a constant if the states are localized. The LSS also provides
a simple way of distinguishing between extended and localized states:
for extended states, level repulsion is expected and the spacings
between energy levels assume a Wigner distribution with variance $(\sigma/\langle s\rangle)^{2}=0.178$
(for the case of the unitary class to which the HFKM belongs), where
$\langle s\rangle$ is the average value of the distribution of level
spacings $s$; for localized states, the level spacing distribution
acquires a Poisson-like shape with a larger variance~\citep{Prodan2011}.
In what follows, we describe the properties of the different phases
in Fig.\,\ref{fig:PhaseDiagram_HFKM_Main}.

\begin{figure}
\centering{}\includegraphics[width=1\columnwidth]{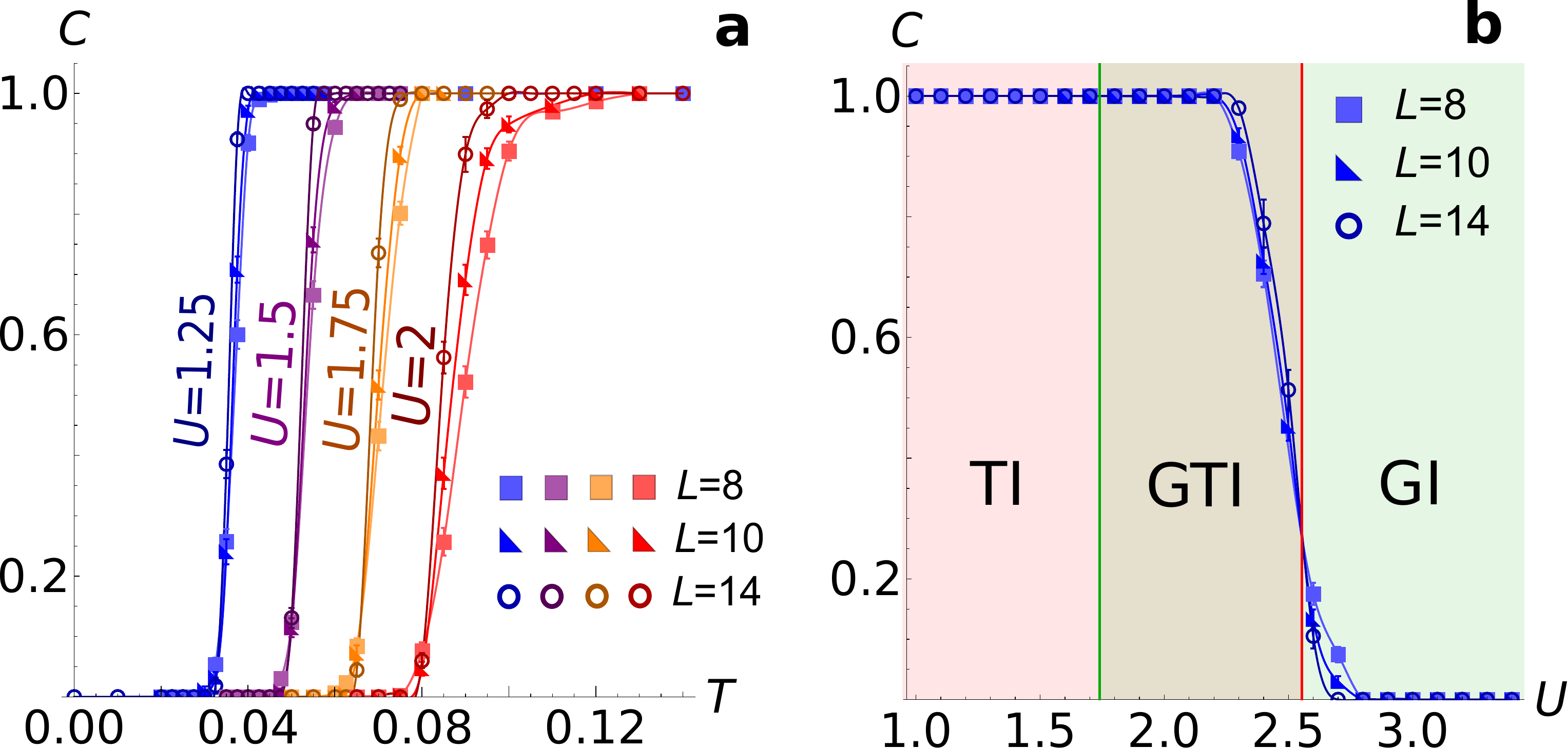}\caption{Chern number computed through averages on Monte Carlo configurations
of $f$-electrons for different system sizes, with fixed $U$ (\textbf{a})
and $T$ (\textbf{b}). The crossing points of these curves were used
to obtain the topological phase transition curve in Fig.\,\ref{fig:PhaseDiagram_HFKM_Main}.
\label{fig:Chern_panel}}
\end{figure}

\paragraph{(CDW).---}

Below the $T_{{\rm {\rm CDW}}}$ curve, dashed-dotted (blue) line
in Figs.\,\ref{fig:PhaseDiagram_HFKM_Main} and \ref{fig:CDW_DOS_panel}\,\textbf{a},
the $f$-electrons start ordering in a checkerboard-like pattern for
which only one of the sublattices is occupied as sketched in the honeycomb
cell in Fig.\,\ref{fig:CDW_DOS_panel}\,\textbf{a}. To better understand
the behavior of the CDW phase transition curve, we perform a mapping
to the 2D antiferromagnetic Ising model at small and large $U$ and
the phase transition curves can be obtained with a perturbative analysis~\citep{SM}.
These curves were computed up to second order in the perturbation
which involves either the terms with $U$ or with the hoppings $t$
and $t_{2}$ and are shown in Fig.\,\ref{fig:CDW_DOS_panel}\,\textbf{a},
as full (red) lines. For $U\ll1$, the $T_{{\rm {\rm CDW}}}(U)$ curve
is quadratic whereas at large interactions, it is inversely proportional
to $U$ -- the agreement with MC is remarkable.

Besides the expected trivial gapped CDW phase (MI-CDW), as found in
Ref.\,\citep{Antipov2016} for the FKM, the HFKM additionally hosts
a topological insulating phase with charge ordering (TI-CDW) along
with a peculiar region of the phase diagram for which the $c$-electron
spectrum is gapless inside the CDW phase (GI-CDW). The former had
already been noticed in Refs.~\citep{Nguyen2013, Zdulski2015} and
contrasts with the results of the spinless fermion Haldane model with
NN interactions for which there is no region of coexistence between
the CDW and TI phases \citep{Varney2010}. Figure.~\ref{fig:CDW_DOS_panel}\,\textbf{b}
compares the DOS inside the MI-CDW and GI-CDW phases for $U=2.5$
and $L=16$, for which the transition between gapped and gapless regimes
can be clearly seen. To ensure that the GI-CDW phase does not stem
from a finite-size effect, we compute the DOS at the Fermi energy
($E=0$) in an energy window corresponding to $1\%$ of the total
bandwidth and by counting the number of states inside, while decreasing
the window's width proportionally to the system size. An example of
this scaling is shown in figure Fig.\,\ref{fig:CDW_DOS_panel}\,\textbf{e}
for a point inside the GI-CDW phase, for which it can be seen that
the DOS($E=0$) is stabilized and does not scale to zero.

\paragraph{(TI and GTI).---}

The TI is a gapped topological phase, i.e., DOS$(E)=0$ for $|E|<\Delta_{\mathrm{Top}}/2$
and Chern number $C=1$, with $\Delta_{\mathrm{Top}}$ being the topological
gap. At $T=0$, the $f$-electrons only occupy one of the sublattices
and therefore act as a staggered potential for the $c$-electrons.
This means that the topological insulating phase exists between $U=0$
and $U=6\sqrt{3}t_{2}\approx1$, value at which the gap closes and
reopens signaling the topological phase transition (TPT) \citep{Haldane1988}.
When we increase $T$, the topological phase still exists and extends
to larger $U$. This is expected for the Haldane model with binary
disorder -- the large $T$ limit, depicted by the vertical lines
in the phase diagram (Fig.\,\ref{fig:PhaseDiagram_HFKM_Main}) at
$T\gg1$ \citep{Goncalves2018} -- where the topological phase is
only destroyed for $U\approx2.7$, meaning there must be a $T$-driven
TPT for $1\lessapprox U\lessapprox2.7$. The corresponding phase transition
curve is shown in Fig.\,\ref{fig:PhaseDiagram_HFKM_Main} as a thin
continuous (red) line \footnote{We notice that the small $U$ behavior of this curve has been qualitatively
predicted with mean-field in Ref.\,\citep{Zdulski2015}.} and some of the Chern number curves used to compile it are shown
in Fig.\,\ref{fig:Chern_panel}. We argue that this $T$-driven topological
transition is reminiscent of the TAI phenomenon, for which disorder-induced
transitions into topological phases can occur. Although there is no
quenched disorder in the system, thermal fluctuations act as to promote
an \textit{annealed disorder}, i.e., the thermal average on the $f$-electron
configurations acts as a disordered potential to the $c$-electrons.
As a result, the topological phase extends into the gapless region
of the phase diagram for higher $T$.

We notice in Fig.\,\ref{fig:CDW_DOS_panel}\,\textbf{c}, that the
topological gap existing in the TI phase is closed in the GTI phase
as one increases the interactions but the Chern number is unchanged
$(C=1)$, as can be seen in Fig.\,\ref{fig:Chern_panel}\,\textbf{b}.
For the TPT from MI-CDW into the TI phase with increasing $T$, the
gap closes and reopens at the phase transition curve. On the other
hand, the TPT from the GTI into the GI phase (the GI is further analysed
below) is accompanied by the merging of the only two extended states
that exist in the spectrum and carry opposite Chern numbers.

\begin{figure}
\centering{}\includegraphics[width=1\columnwidth]{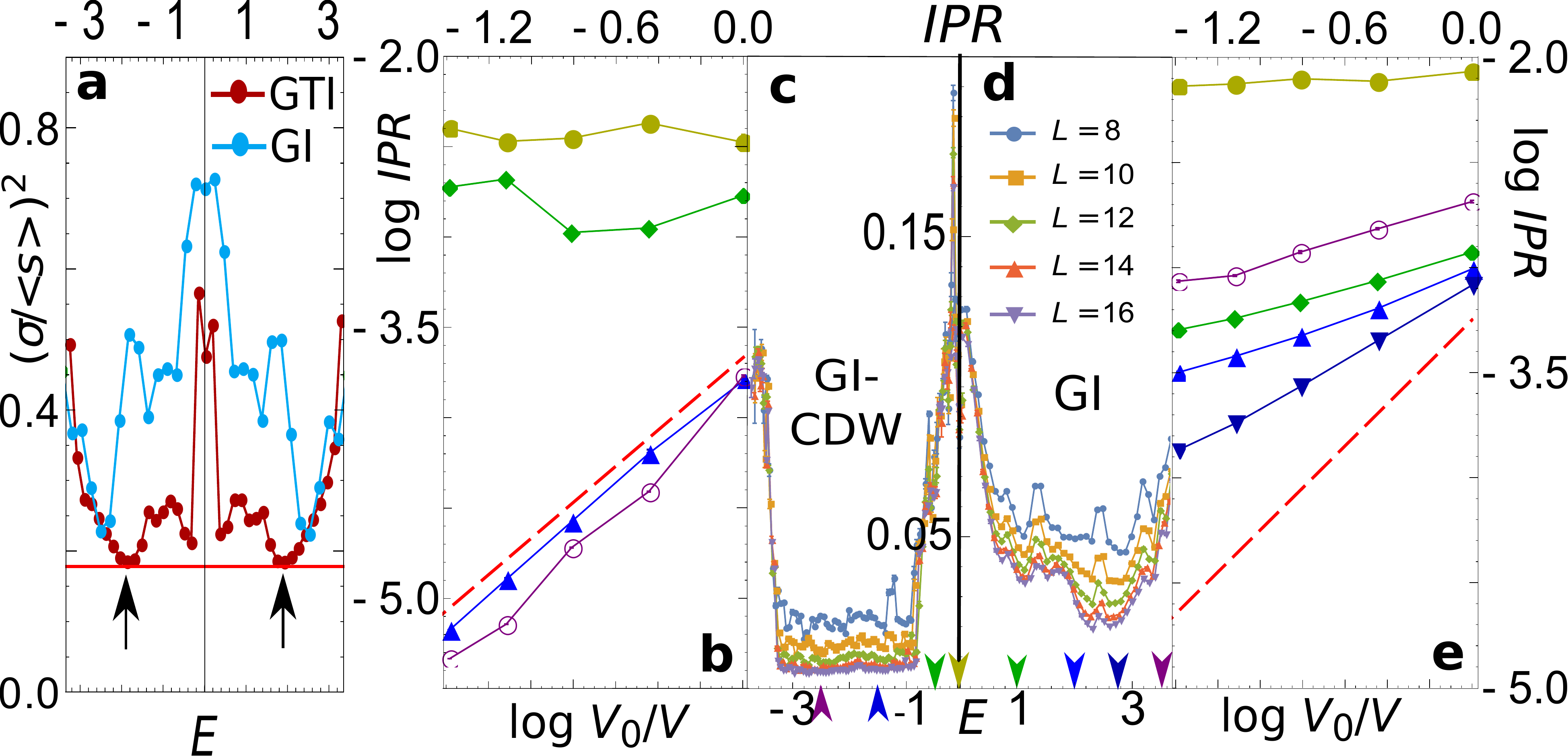}\caption{\textbf{a}, Standard deviations of the LSS distributions obtained
for different energies in the GI {[}GTI{]} phase for $(U,T)=(2,0.1)$
{[}$=(3.5,0.2)${]} and $L=14$. The horizontal (red) line corresponds
to $\sigma/\langle s\rangle=0.178$ which is the standard deviation
of the Wigner distribution associated to extended states. The two
extended states existing in the GTI phase are marked with arrows.
\textbf{b (e)}, Finite size scaling of the IPR with the system volume
$V$ for the energies marked with the arrows in \textbf{c (d)}, that
shows the IPR for different sizes in the GI-CDW {[}GI{]} phase for
$(U,T)=(2.5,0.085)$ {[}$=(3.5,0.2)${]}. The IPR shown in \textbf{c}
(\textbf{d}) for negative (positive) energies is symmetric in $E$.
The red dashed lines shown in \textbf{b},\textbf{e} have a unit slope
and indicate the scaling IPR$\sim V^{-1}$. The colors of the arrows
that select specific energies in \textbf{c} (\textbf{d}) match the
corresponding scaling curves in \textbf{b} (\textbf{e}).\label{fig:localization_panel}}
\end{figure}

Topological phases are robust at finite $T$ provided the thermal
fluctuation energy $k_{B}T$ does not exceed the energy separation
of the extended states, as all the eigenstates in-between are localized
and cannot change the Chern number, similarly to the case of the integer
quantum Hall effect. This condition breaks down near the TPT curve
of Fig.\,\ref{fig:PhaseDiagram_HFKM_Main}. However, as shown in
Fig.\,\ref{fig:localization_panel}\,\textbf{a}, only slightly away
from the TPT line, the extended states already have an energy separation
$\Delta E\approx4\gg k_{B}T\approx0.1$, allowing for a $T$-driven
TPT into gapped and gapless TIs.

\paragraph{(GI and MI).---}

Increasing $U$ from the GTI phase leads to an interaction-driven
TPT into a trivial gapless insulating phase (GI). If we continue increasing
$U$, the $c$-electron spectrum acquires a Mott-like gap (MI). The
corresponding DOS within the GI and MI phases is exemplified in Fig.\,\ref{fig:CDW_DOS_panel}\,\textbf{d}.
This phase transition resembles the one found for the 2D FKM in Ref.\,\citep{Antipov2016},
for increasing interactions between an Anderson and a Mott insulator.

\paragraph{Gapless insulators.---}

We report, in Fig.\,\ref{fig:localization_panel}\,\textbf{a,} the
LSS results for the GTI and GI phases. In the GTI phase, the standard
deviation of the level spacing distributions, $\sigma/\left\langle s\right\rangle $,
has the expected value for the Wigner distribution (horizontal line)
at the particular energies corresponding to the two extended states
that carry opposite Chern numbers. Away from these particular values,
and within all the GI phase, $\sigma/\left\langle s\right\rangle $
raises above the Wigner distribution prediction signaling the localization
of eigenstates. The IPR, for the GI phase, is depicted in Fig.\,\ref{fig:localization_panel}\,\textbf{d}.
Around $E=0$, it is almost unchanged with the system size thus, the
corresponding states are undoubtedly localized. However, for larger
$|E|$ values, the IPR decreases with the system size. A finite-size
analysis is shown in Fig.\,\ref{fig:localization_panel}\,\textbf{e},
where the unit slope associated with the scaling ${\rm IPR}\propto V^{-1}$
is depicted by the (red) dashed line. Nonetheless, the slopes at different
energies are always smaller than one, suggesting that localization
is robust for every energy in the GI phase. These results are compatible
with the following scenario: outside the CDW phase, spatial correlations
between $f$-electron occupations in GTI and GI phases decay exponentially
with a characteristic length $\xi$. For distances larger then $\xi$,
the disorder potential felt by the $c$-electrons becomes uncorrelated.
These phases smoothly extend to large $T$, for which $\xi\simeq1$
and where disordered effects become equivalent to those of a binary
quenched potential \citep{Goncalves2018}.

For the GI-CDW, Figs.\,\ref{fig:localization_panel}\,\textbf{b}
and \ref{fig:localization_panel}\,\textbf{c}, a similar analysis
suggests that although the eigenstates are localized around $E=0$,
there are also regions of extended states. Figure\,\ref{fig:localization_panel}\,\textbf{c}
shows that the IPR becomes smaller with $V$ for $-3\lesssim E\lesssim-1$
(and $1\lesssim E\lesssim3$, not shown) and Fig.\,\ref{fig:localization_panel}\,\textbf{b}
indicates that for two energies in this interval, the IPR indeed scales
with $V^{-1}$ for the used system sizes. This is in apparent contradiction
with results for $\sigma/\left\langle s\right\rangle $ (not shown),
where all energies rise above the Wigner distribution prediction indicating
that all eigenstates should be localized. These seemingly contradicting
facts can be reconciled by noticing that, inside the CDW phase, $\xi$
diverges and the disordered potential experienced by the $c$-electrons
becomes long-range correlated. In two-dimensions, systems with long-range
spatially correlated disorder have been shown to support spectral
regions of extended states \citep{Taras_Semchuk2001, Hilke2003, DeMoura2007},
moreover, Wigner distribution predictions are expected not to hold
for such kind of disorder.

Figure~\ref{fig:PhaseDiagram_HFKM_Main} shows that the GI-CDW phase
is created, starting from the $T=0$ gapped CDW, by increasing $T$.
Our results show that the gap starts being populated by localized
states induced by the thermal fluctuations. Here, again, disorder
is correlated and may support extended states for a finite disorder
strength. The important question is whether a region of extended states
still survives upon entering the GI-CDW phase or if all states are
already localized for this value of $T$. Although our results strongly
suggest the former, we cannot definitely exclude the latter scenario
which will require working with substantially larger systems sizes.
If confirmed, the coexistence of spectral regions of extended and
localized states would correspond to one of the first examples of
a many-body mobility edge in a strong interacting system and may suggest
similar phenomena to be present in the case of finite mass-ratio between
electronic species.

Summarizing the central result of our work: we introduce the HFKM
model allowing to effectively study the interplay of topology and
interactions at finite temperatures and provide a complete characterization
of the phase diagram. We show the possibility of having a temperature-driven
topological phase transition into gapped and gapless topological insulators;
finally, we find an insulating charge ordered state with gapless excitations
where spectral regions of extended and localized states seem to coexist
due to the long range nature of the interaction-induced disorder potential.

All the ingredients for the experimental realization of the HFKM with
ultracold atoms in optical lattices are separately available: there
are recent implementations of mass unbalanced fermions \citep{Jotzu2015,Greif2015};
and, the Haldane model has recently been successfully realized \citep{esslingerHaldane}.
A direct verification of our results should therefore be achievable
with state-of-the-art technology. 
\begin{acknowledgments}
The M.G., P.R. and E.V.C acknowledge partial support from FCT-Portugal
through Grant No. UID/CTM/04540/2013. M.G. and P.R. acknowledge support
by FCT-Portugal through the Investigador FCT contract IF/00347/2014.
R.M. acknowledges support of the NSFC Grants No. 11674021 and No.
11650110441, and NSAF-U1530401. The computations were performed in
the Tianhe-2JK at the Beijing Computational Science Research Center
(CSRC). 
\end{acknowledgments}

 \bibliographystyle{apsrev4-1}
\bibliography{references}

\clearpage\onecolumngrid 
\begin{center}
\textbf{\large{}{}Supplementary Materials: }{\large{}}\\
{\large{} }\textbf{\large{}{} Temperature-driven gapless topological
insulators}{\large{}}\\
{\large{} }{\large\par}
\par\end{center}

\begin{center}
\vspace{0.3cm}
 
\par\end{center}

\vspace{0.6cm}

\twocolumngrid

\beginsupplement

\section*{Perturbative analysis}

\label{sec:perturbation_theory}

By making a perturbative expansion for the effective Hamiltonian $\mathcal{H}(\{n_{f}\})$
defined in Eq.\,(\ref{eq:effective_hamil}) it is possible to study
the small and large $U$ regions of the phase diagram. This expansion
allows us to write the Hamiltonian of the HFKM in the form of an effective
2D antiferromagnetic Ising model that only depends on the $f$-electrons'
degrees of freedom. In this way, we can see the CDW phase transition
curves of the effective Ising models as an approximation of the exact
phase transition curve of the HFKM on the limits of concern.

We start by defining the Hamiltonian matrix $\bm{H}$

\begin{equation}
\bm{H}=-t\sum_{\langle i,j\rangle}\ket{i}\bra{j}+{\rm i}t_{2}\sum_{\langle\langle i,j\rangle\rangle}\nu_{ij}\ket{i}\bra{j}+\frac{U}{2}\sum_{i}s_{i}\ket{i}\bra{i},\label{eq:Hamiltonian_matrix}
\end{equation}

\noindent where we introduced the Ising variables $s_{i}=2n_{i}^{f}-1=\pm1$.
The propagator of $c$-electrons is then simply given by $\bm{G}=({\rm i}\omega_{n}-\bm{H})^{-1}$.
Once the HFKM Hamiltonian is quadratic in the $c$-electron's fields
for a given configuration $\{n_{f}\}$, the formalism of Gaussian
path integrals can be employed to write \citep{coleman2015introduction}

\begin{equation}
\mathcal{H}=-\frac{1}{\beta}{\rm Tr}_{c}\ln(-\bm{G}^{-1})-\frac{U}{2}\sum_{i}n_{i}^{f},\label{eq:Heff_Green}
\end{equation}

\noindent where the trace is taken over the fermionic degrees of freedom
and was extended to incorporate the sum in the Matsubara frequencies.
If we separate $\bm{H}$ in the unperturbed and perturbed terms, respectively
$\bm{H}_{0}$ and $\bm{H}_{1}$, we can show that

\begin{equation}
\mathcal{H}=\mathcal{H}_{0}-\frac{U}{2}\sum_{i}n_{i}^{f}+\frac{1}{\beta}\sum_{k=1}^{+\infty}\frac{1}{k!}{\rm Tr}\Big[(\bm{G}_{0}\bm{H}_{1})^{k}\Big],\label{eq:perturbation_expansion}
\end{equation}

\noindent where $\bm{G}_{0}^{-1}={\rm i}\omega_{n}-\bm{H}_{0}$ and
$\mathcal{H}_{0}=-\frac{1}{\beta}{\rm Tr}\ln(-\bm{G}_{0}^{-1})$.
Equation\,(\ref{eq:perturbation_expansion}) provides a useful starting
point for our perturbative analysis and can be applied for the small
and large $U$ limits. For small $U$, we have that $\bm{H}_{0}$
and $\bm{H}_{1}$ contain respectively the hopping and $U$ dependent
terms in Eq.\,(\ref{eq:Hamiltonian_matrix}), while for large $U$,
they interchange. For the perturbative analysis, the expansion is
made up to second order in $\bm{H}_{1}$. For small $U$, the effective
Ising Hamiltonian $\mathcal{H}_{SU}$ is

\begin{equation}
\mathcal{H}_{SU}=U^{2}\sum_{i,j}J_{ij}(R)s_{i}s_{j},
\end{equation}

\noindent where the sum is over pairs of neighbors and

{\small{}{}{} 
\begin{equation}
J_{ij}(R)=\begin{cases}
\int_{-\infty}^{+\infty}d\omega\text{ }\frac{A_{\hexagon}^{2}}{16\pi^{3}v^{4}}(m^{2}-\omega^{2})K_{0}^{2}\Big(\frac{R}{v}\sqrt{\omega^{2}+m^{2}}\Big) & \text{, (i) }\\
\int_{-\infty}^{+\infty}d\omega\frac{A_{\hexagon}^{2}}{16\pi^{3}v^{4}}(m^{2}+\omega^{2})K_{1}^{2}\Big(\frac{R}{v}\sqrt{\omega^{2}+m^{2}}\Big) & \text{, (ii) }
\end{cases}\label{eq:couplings_smallU}
\end{equation}
}{\small\par}

\noindent In the above expression, $R$ is the absolute distance between sites $i$ and $j$, $m=3\sqrt{3}t_2$ is the Haldane model's topological gap, $A_{\hexagon}=3\sqrt{3}a^2/2$ is the area of the honeycomb unit cell (with $a$ being the lattice constant) and $v=3t/2$ is the Fermi velocity. Finally, $K_{0}$ and $K_{1}$ are modified Bessel functions
of the second kind. Conditions (i) and (ii) correspond respectively to $i$ and $j$ in
the same and in different sublattices.

On the other hand, the effective
Ising Hamiltonian $\mathcal{H}_{LU}$ for large $U$ is

\begin{equation}
\mathcal{H}_{LU}=\sum_{\langle i,j\rangle}\frac{t^{2}}{2U}s_{i}s_{j}+\sum_{\langle\langle i,j\rangle\rangle}\frac{t_{2}^{2}}{2U}s_{i}s_{j}
\end{equation}

Based on the effective Ising models, the critical temperature curves
were estimated and are shown in Fig.\,\ref{fig:CDW_DOS_panel} along
with the numerical results. For small $U$, the critical temperature
was estimated up to next-NNN and NNN were not considered as they can
be neglected for the studied case of $t_{2}=0.1t$. For large $U$,
the second term in $\mathcal{H}_{LU}$ can also be neglected for the
case of interest. 
\end{document}